\documentclass[aps,pra,twocolumn,groupedaddress,showpacs]{revtex4}

\def\proof{{\em Proof:  }}
\newcommand{\QED}{\hspace*{\fill}\mbox{\rule[0pt]{1.5ex}{1.5ex}}}
\def\tphi{\tilde{\phi}}

\def\openone{\leavevmode\hbox{\small1\kern-3.8pt\normalsize1}}
\def\RR{{\rm I\kern-.2emR}}
\def\extr{{\rm extr}}
\def\tr{{\rm tr}\; }

\newcommand{\ket}[1]{| #1 \rangle}
\newcommand{\bra}[1]{\langle #1 |}
\newcommand{\proj}[1]{\ket{#1}\! \bra{#1}}

\newcommand{\bitem}{\begin{itemize}}
\newcommand{\eitem}{\end{itemize}}
\newcommand{\benum}{\begin{enumerate}}
\newcommand{\eenum}{\end{enumerate}}
\newcommand{\beq}{\begin{equation}}
\newcommand{\eeq}{\end{equation}}
\newcommand{\beqa}{\begin{eqnarray}}
\newcommand{\eeqa}{\end{eqnarray}}
\newtheorem{definition}{Definition}

\newtheorem{theorem}{Theorem}
\newtheorem{example}{Example}
\newtheorem{proposition}{Proposition}

\newtheorem{lemma}{Lemma}
\newtheorem{corollary}{Corollary}
\newtheorem{conjecture}{Conjecture}

\newtheorem{remark}{Remark}
\newcommand{\bproof}{\begin{proof}}
\newcommand{\eproof}{\end{proof}}
\newcommand{\bprop}{\begin{proposition}}

\newcommand{\bdef}{\begin{definition}}

\def\C{{\bf C}}
\def\R{{\bf R}}

\begin{document}


\title{Separable balls around the maximally mixed multipartite quantum states}


\author{Leonid Gurvits and Howard Barnum}
\affiliation{ CCS-3, Mail Stop B256, Los Alamos National Laboratory,
Los Alamos, NM 87545} 
\affiliation{}


\date{February 12, 2003}

\begin{abstract}
We show that for an $m$-partite quantum system, there is 
a ball of radius $2^{-(m/2-1)}$ in Frobenius norm,
centered at the identity matrix,   
of separable (unentangled) positive semidefinite
matrices.  This can be used to derive an $\epsilon$ below 
which mixtures of $\epsilon$ of 
any density matrix with $1 - \epsilon$ of
the maximally mixed
state will be separable.  The $\epsilon$ thus obtained is
exponentially better (in the number of systems) than existing
results.  This gives a number of qubits below which
NMR with standard pseudopure-state preparation techniques
can access only unentangled states; with parameters realistic
for current experiments, this is 23 qubits (compared to 13 qubits
via earlier results).  A ball of radius $1$ is obtained for
multipartite states separable over the reals.
\end{abstract}


\maketitle

\newcommand{\thm}{\begin{theorem}}
\newcommand{\lem}{\begin{lemma}}
\newcommand{\pro}{\begin{proposition}}
\newcommand{\dfn}{\begin{definition}}
\newcommand{\rem}{\begin{remark}}
\newcommand{\xam}{\begin{example}}
\newcommand{\cnj}{\begin{conjecture}}
\newcommand{\que}{\begin{question}}
\newcommand{\cor}{\begin{corollary}}
\newcommand{\prf}{\noindent{\bf Proof:} }
\newcommand{\ethm}{\end{theorem}}
\newcommand{\elem}{\end{lemma}}
\newcommand{\epro}{\end{proposition}}
\newcommand{\edfn}{\bbox\end{definition}}
\newcommand{\erem}{\bbox\end{remark}}
\newcommand{\exam}{\bbox\end{example}}
\newcommand{\ecnj}{\bbox\end{conjecture}}
\newcommand{\eque}{\bbox\end{question}}
\newcommand{\ecor}{\end{corollary}}

\def\sup{^}
\def\Tp{Tchebyshef polynomial}
\def\Tps{TchebysDeto be the maximafine $A(n,d)$ l size of a code with distance 
$d$hef polynomials}
\newcommand{\rarrow}{\rightarrow}
\newcommand{\larrow}{\leftarrow}
\newcommand{\grad}{\bigtriangledown}

\overfullrule=0pt
\def\setof#1{\lbrace #1 \rbrace}
\section{Introduction}
Entanglement is an important quantum resource, useful in 
quantum computation, cryptography, and communication protocols.
Entangled quantum states are those that cannot be expressed
as a mixture of product states.  That is, if $\rho$ is an
entangled state of $m$ systems (an ``$m$-partite state''),  
there is no way to choose
probabilities $p_i$ and states $\rho_i^1,...,\rho_i^m$ for
systems $1$ through $m$, such that
\beqa \label{definition of separable}
\rho = \sum_i p_i \rho_i^1 \otimes \cdots \otimes \rho_i^m\;.
\eeqa
In this paper, we provide a simple geometric condition
sufficient to guarantee separability (non-entanglement)
of an $m$-partite
state:  that the state is proportional to the identity 
matrix plus a Hermitian perturbation $\Delta$ whose
Frobenius norm ($2$-norm) is no greater than 
$2^{-(m/2 -1)}$.  This is exponentially better than 
the best previous bounds we are aware of 
\cite{Braunstein99a, Rungta2001a}.
It can be used to obtain 
balls of normalized  separable states.
Because the set of separable (unentangled)
density matrices is convex, and the size of the largest
ball that fits inside it (as well as the smallest ball that
covers it) is important to complexity-theoretic questions 
involving convex sets, we expect the result to have applications
in complexity questions about entanglement, such as the
complexity of deciding whether or not a 
multipartite state is entangled.  Equally importantly, it can 
help determine whether or not entanglement is present in interesting
theoretical and experimental situations.  For example, though
the utility of the criterion is emphatically not restricted
to such states, it gives us a bound on the ``polarization''
$\epsilon$ below which ``pseudopure'' states of the form
$(1 - \epsilon)I/d + \epsilon \pi$, with $\pi$ pure, $I$ 
the identity operator on a multipartite state-space of
overall dimension $d$ are separable.  Applied to a standard
nuclear magnetic resonance
(NMR) quantum information processing
(QIP) protocol, pseudopure-state liquid NMR, at 300 Kelvin and 
a 11 Tesla external field, it tells us that
unless we have 23 or more nuclear-spin qubits, only separable
states can be produced (compared to 13 qubits 
using the bound in \cite{Braunstein99a}).

The methods of \cite{Braunstein99a, Rungta2001a}
are very different from ours: they 
expand the density matrix in an 
overcomplete basis of pure states, and find
conditions guaranteeing positivity of all coefficients
of the expansion, thus giving an explicit decomposition
of the form (\ref{definition of separable}).
In contrast, the methods we use  have a 
nonconstructive flavor:
although they establish that any $m$-partite
(unnormalized) density matrix within a distance
$1/2^{(m/2-1)}$ is separable, they do not provide
an explicit ``separable representation'' (\ref{definition of separable})
of it.   Our methods use general concepts of matrix
theory and convex analysis in terms of which the problem
is naturally formulated,  and 
involve only short and elementary calculations.
This paper is a natural sequel to  \cite{GB02}, in
the sense that almost the same mathematics is used.
The main (though quite simple) novelty here is
a generalization of {\bf separability}, so-called
$(C_{1} \otimes C_{2} \otimes ... \otimes C_{m})$-separability ,
where $C_{i}$ are matrix cones (see Definition 2 below).
This generalization arises naturally in extending the
bipartite result of \cite{GB02}
to $m$-partite systems with $m \geq 3$.
A reader comfortable with the technique used in \cite{GB02}
should have no extra problems in understanding this paper.

\section{Notation and mathematical preliminaries}

The mathematical notion of a ``regular''
positive cone (which we will just call cone)
is basic
in quantum information science, especially in the study of
entanglement.  This is so  because the unnormalized 
quantum states, the unnormalized separable
states of a multipartite quantum system, the completely positive maps,
the positive maps, and many other sets of interest form such cones.  
(Appropriate normalization-like
conditions, such as unit trace for states, or trace-preservation or
trace-nonincrease for maps, are usually just additional linear
equalities or inequalities.)
In this section, we review regular positive cones and related
notions; 
background and
preliminaries specific to the separable cones (of unentangled states)
appear at the beginning of the next section.

\begin{definition}
A {\em positive cone} is
a subset $K$ of a real vector space $V$ closed under multiplication by
positive scalars.
It is called {\em regular} if it is (a) convex (equivalently,
closed under addition:  $K + K = K$), 
(b)
 generating ($K-K=V$, equivalently $K$
linearly generates $V$,) 
(c) 
pointed ($K \cap -K = \emptyset$, so that it
contains no nonnull subspace of $V$), and (d) 
topologically closed (in the
Euclidean metric topology, for finite dimension).  
\end{definition}

Such a positive cone induces a  partial order
$\ge_K$ on $V$, defined by $x \ge_K y := x - y \in K$. 
It is ``linear-compatible'':
inequalities can be added, and multiplied by positive scalars.  
A set $S$ is said to {\em generate} a cone
$K$ if $K$ is the set of {\em positive} linear combinations of
elements of $S$.
The topological closure condition guarantees
that such a cone is generated (via addition) 
by its {\em extreme rays}.
These are sets $R_x := \{\lambda x: \lambda \ge 0 \}$ such
that no 
$y \in R_x$ can be written as a convex combination of 
elements of $C$ that are not in $R_x$.  We will not make 
much use of closure and extremality, 
but at some points we use the fact 
that positive semidefinite (PSD) matrices can be written as
convex combinations of rank-one PSD matrices (these being
the members of the extreme rays of the cone of PSD matrices).

Duality is often a useful tool when dealing with cones.  The dual
vector space $V^*$ for real $V$ is the space of linear functions
(``functionals'') from $V$ to $\R$; the dual cone $C^* \subset V^*$ of
the cone $C \subset V$ is the set of such linear functionals which are
nonnegative on $C$.  For finite dimensional vector spaces, 
$V^*$ is isomorphic to $V$ as a vector space (they
have the same dimension).  However, this isomorphism is not 
canonical; any 
nonsingular linear maps from $V$ onto $V^*$ does the job.  
For a slight improvement in clarity below,
for $\alpha \in V^*$, $x \in V$, we write the value of $\alpha$ 
on $x$ as $\alpha[x]$, rather than $\alpha(x)$.

We define the adjoint $\phi^\dagger: V_2^* \rightarrow V_1^*$ 
of a linear map $\phi: V_1 \rightarrow 
V_2$ via 
\beqa
\phi^\dagger(\alpha)[x] = \alpha[\phi(x)]\;,
\eeqa
for all $\alpha \in V^*, x \in V$.  

An inner product, written $\langle , \rangle$, 
on $V$ distinguishes a particular isomorphism $\zeta$ between
$V$ and $V^*$, defined by requiring that $\zeta(x) \in V^*$ 
satisfy $\zeta(x)[y] = \langle x, y \rangle$, for all $y \in V$.
From now on, we will assume
such a choice of inner product (there will be a natural choice in our 
applications), and canonically identify
$V$ with $V^*$ via $\zeta$.  Thus we will view the dual cone
$C^*$ as $\{ y \in V: \langle y, x \rangle \ge0, \forall x \in C\}$, 
and the adjoint of $\phi: V_1 \rightarrow V_2$ as 
$\phi^\dagger: V_2 \rightarrow V_1$ defined via
\beqa
\langle B, \phi(A) \rangle = \langle \phi^\dagger(B), A \rangle \;,
\eeqa
for all $A \in V_1, B \in V_2$.

We say a linear map $\phi: V_1 \rightarrow V_2$ is $C_1$-to-$C_2$
positive, for cones $C_1 \subset V_1$, $C_2 \subset V_2$, if
$\phi(C_1) \subseteq C_2$.  Under either definition of $\phi^\dagger$,
the following proposition is easily (but instructively) verified.
\begin{proposition}
If $\phi(C_1) \subseteq C_2$ then
$\phi^\dagger(C_2^*) \subseteq C_1^*$.
\end{proposition}

The {\em positive semidefinite cone} $PSD(d)$ in the real
linear space of Hermitian $d \times d$ matrices, is the set
of matrices $M$ such that  $x^\dagger M x \ge 0$ for all 
$x \in \C^d$.  It is self-dual;  if we use the
trace inner product 
$\langle X, Y \rangle := \tr X Y$ to identify $V^*$ with $V$, 
it is not only isomorphic to, but equal to, its dual.
We will denote by ``$\succeq$'' the partial order induced
by this cone, and often write $M \succeq 0$ for the equivalent
$M \in PSD(d)$.  

We will have several occasions to use the following proposition,
which follows from the fact 
that for normal (including
Hermitian) matrices,
$\Delta$, $||\Delta||_\infty$ is the largest 
modulus of an eigenvalue of $\Delta$).  
\begin{proposition} \label{elementary, my dear Watson}
Let $\Delta$ be Hermitian.  Then $I + \Delta \succeq 0$ is 
equivalent to $||\Delta||_\infty \le 1$.  
\end{proposition}

\section{Separable cones}
Let us consider an $m$-partite unnormalized density matrix (i.e. just
positive semidefinite)
$$
\rho: H_{1} \otimes H_{2} \otimes ... \otimes H_{m} \longrightarrow
H_{1} \otimes H_{2} \otimes ... \otimes H_{m}
$$
Let $dim(H_{i}) = d_{i}, 1 \leq i \leq m$. Then any such 
$$
\rho = \{\rho(i_{1},i_{2},...,i_{m}; j_{1},j_{2},...,j_{m}) \}, 1 \leq
i_{k}, j_{k} \leq d_{k}; 1 \leq k \leq m \}.
$$
Let us block-partition $\rho$ with respect to the first index :
\beq
\label{block}
\rho := 
\left( \begin{array}{cccc}
		  \rho^{1,1} &\rho^{1,2} & \dots & \rho^{1,d_{1}}\\
		  \rho^{2,1} &\rho^{1,2}  & \dots & \rho^{2,d_{1}}\\
		  \dots &\dots & \dots & \dots \\
		  \rho^{d_{1},1} &\rho^{d_{1},2}  & \dots & \rho^{d_{1},d_{1}}
		  
\end{array} \right).
\eeq
where the blocks 
$$
\rho^{i,j}: H_{2} \otimes ... \otimes H_{m} \longrightarrow H_{2} \otimes ... \otimes H_{m}
$$
and
$$
\rho^{i,j} = \{\rho(i,i_{2},...,i_{m}; j,j_{2},...,j_{m} \},   1 \leq  i_{k}, j_{k} \leq d_{k};   2 \leq k \leq m \}.
$$
\dfn
The linear space of $N \times N$ complex matrices is denoted  $M(N)$, 
the linear space over the reals of  
$N \times N$ real matrices is denoted $Mat(N)$ ,
the linear space of real symmetric $N \times N$ matrices is denoted $RSym(N)$, the linear space over reals of 
$N \times N$ complex hermitian matrices is denoted as $Her(N)$.  The
space of complex block matrices, $K$ blocks by $K$ blocks, with blocks 
in $M(N)$, is denoted $Block(K,N)$\\
Consider cones $C_{i} \subset M(d_{i}),  1 \leq i \leq m$. A matrix
$$
\rho: H_{1} \otimes H_{2} \otimes ... \otimes H_{m} \longrightarrow H_{1} \otimes H_{2} \otimes ... \otimes H_{m}
$$
(i.e. $ \rho \in M(d_{1} d_{2} ... d_{m}) $ )  is
called $(C_{1} \otimes C_{2} \otimes ... \otimes C_{m})$-separable if it
belongs to the cone generated by the set $\{ A_{1}  \otimes A_{2} \otimes ... \otimes A_{m}: A_{i} \in C_{i} , 1 \leq i \leq m \}$ .
\end{definition}

This is trivially equivalent to the recursive definition:
$Sep(C_1, C_2,..., C_m)$ is the cone generated by the pairs 
$A_1 \otimes B$ with $A_i \in C_1, B \in Sep(C_2,...,C_m)$.

\noindent
{\bf Examples:}
\noindent
1.
Let for all $1 \leq i \leq m$ the cone $C_{i}$ be the cone of positive
semidefinite matrices, denoted by $PSD(d_{i})$. In this case the
definition of $(C_{1} \otimes C_{2} \otimes ... \otimes
C_{m})$-separability is equivalent to the standard notion of
separability of multiparty unnormalized density matrices. We will denote the
corresponding cone of separable multiparty unnormalized density matrices as
$Sep(d_1,d_2,...,d_m)$. \\
\noindent 2.
Let for all $1 \leq i \leq m$ the cone $C_{i}$ be the cone of 
positive semidefinite matrices with real entries.  We call the  
the corresponding cone
the cone of real-separable multiparty density matrices and denote it by
$RSep(d_1,d_2,...,d_m)$. 

We now recursively define a subspace $RLin(d_1,...,d_m),$
which we show is the minimal linear subspace
(over the reals) of the symmetric matrices 
$RSym(d_1,...,d_m)$  that 
contains the real separable cone $RSep(d_1,...,d_m)$.

\begin{definition}
\label{def: RLin}
$\rho \in
RLin(d_1,d_2,...,d_m)$  iff in the block-partition \ref{block},
$\rho^{i,j} \in RLin(d_2,...,d_m)$ and $\rho^{i,j} = \rho^{j,i} ( 1
\leq i,j \leq d_1 )$;  RLin(d) = RSym(d).
\end{definition}

It
is easy to prove that $\rho: H_{1} \otimes H_{2} \otimes ... \otimes
H_{m} \longrightarrow H_{1} \otimes H_{2} \otimes ... \otimes H_{m}$
is real-separable iff $\rho$ is separable and $\rho \in
RLin(d_1,d_2,...,d_m)$. 
In fact, we will show 
\begin{proposition} \label{prop: RLin and RSep}
$RLin(d_1,d_2,...,d_m)$ is the minimal
linear subspace (over the reals) of $RSym(d_1 d_{2}...d_m)$ which contains
$RSep(d_1,d_2,...,d_m)$. 
\end{proposition}

\noindent
{\bf Proof:}
It is  
clear that $RLin(d_1....d_m)$ is a subspace of (``$\le$'') 
$RSym(d_1...d_m)$.
To be explicit, symmetry means $\rho(i_1, ..., i_m, j_1,...,j_m) = 
\rho(j_1, ..., j_m, i_1,...,i_m)$; this follows from the definition 
of $RLin$ and $RLin(d_2...d_m)$'s being a subspace of 
$RSym(d_2,...,d_m)$.  That establishes our induction step;
the base case $RLin(d_m) = RSym(d_m)$ clearly also holds.

Suppose $X \in RSep(d_1,...,d_{m-1})$.  That is,  
$X = \sum_k A_k \otimes B_k$, $A_k 
\in RSep(d_1, ..., d_{m-1}), B_k \in RSym(d_m)$.    
By the induction hypothesis,
$A_k \in RLin(d_1,...,d_{m-1})$.
Block-partitioning with
respect to the second system, 
\beqa
X^{ij} = \sum_k B_k(i,j) A_k = \sum_k B_k(j,i) A_k = X^{ji}\;,
\eeqa
These blocks are in $RLin(d_1...d_{m-1})$ because $A_k$ are;
consequently $X \in RLin(d_1,...,d_m)$.  The base case
is trivial: $RSep(d_1) \subseteq (RSep(d_1) \cap RLin(d_1))$
holds with equality because $RSep(d_1)  \equiv PSD(d_1)$ 
and $RLin(d_1) = RSym(d_1)$.  

For the opposite direction, let 
$X \in RLin(d_1,...d_m), Sep(d_1,...,d_m)$.  
By separability, 
\beqa \label{yabba dabba doo}
X = \sum_k A_k \otimes B_k \otimes \cdots \otimes Z_k\;,
\eeqa
where $A_k \in PSD(d_1), B_k \in PSD(d_2), Z_k \in PSD(d_m)$
(and no restriction to $m=26$ is intended!).
Let $A_k = A^1_k + i A^2_k$
with $A^1_k$ real symmetric, $A^2_k$ real skew-symmetric,
and similarly for $B$.  Substituting these in 
(\ref{yabba dabba doo}) and keeping only terms with
an even number of imaginary factors (since 
$X \in RLin$), and block-partitioning 
the matrix according to the first subsystem,
each block has the form:
\beqa
\sum_k A_k^1(i,j) R_k + \sum_k A_k^2(i,j)  S_k\;,
\eeqa
Thus $X = X_1 + X_2$, where
the  first term is block-symmetric, the second block-skew-symmetric.
This second term must therefore be zero.  By the recursive
definition of $RLin$ (and $Sep$) we have that, for each fixed value
of $i,j$, $X^{ij}$ must be block-symmetric 
when partitioned according
to the second (``B'') system.  This block is
\beqa
\sum_k A_k^1(i,j) B^1_k(m,n)  C^k & \otimes & \cdots Z^k \nonumber \\
+ \sum_k A_k^1(i,j) B^2_k(m,n) C^k & \otimes & \cdots Z^k
\eeqa
and again only the first component is nonzero.  Proceeding thus
through all the subsystems, all terms with a skew-symmetric factor
must be zero and we have:
\beqa
X = A_k^1 \otimes B_k^1 \otimes \cdots Z_k^1\;,
\eeqa with each of 
$A_k, B_k, \dots Z_k$ real symmetric and positive semidefinite, 
i.e. $X \in RSep$.

\QED

The next lemma gives a simple but very useful criterion for
$PSD(d_{1}) \otimes C(d_{2})$-separability for any cone
$C(d_{2})$ of Hermitian matrices.  It is a mild generalization of 
the necessary and sufficient
criterion (cf. \cite{Woronowicz76a}, \cite{Horodecki96a}) for
ordinary ($C(d_2) = PSD(d_2)$) bipartite separability, 
that every positive linear map, applied to one subsystem
of the bipartite system (i.e. to every block of its block
density matrix) gives a positive semidefinite matrix.  

\dfn A linear operator $\phi: M(d_{2})
\longrightarrow M(N)$ is called $C(d_{2})$-positive if $\phi(C(d_{2}))
\subset PSD(N)$.  If $X$ is a block matrix as in (\ref{block}),
$X_{i,j} \in M(d_{2})$ and $\phi: M(d_{2}) \longrightarrow M(N)$ is a
linear operator then we define 
\beq \tilde{\phi}(X) := \left(
\begin{array}{cccc} \phi(X^{1,1}) & \phi(X^{1,2}) & \dots &
\phi(X^{1,d_{1}})\\ \phi(X^{2,1}) & \phi(X^{2,2}) & \dots &
\phi(X^{2,d_{1}})\\ \dots &\dots & \dots & \dots \\ \phi(X^{d_{1},1})
& \phi(X^{d_{1},2}) & \dots & \phi(X^{d_{1} ,d_{1}})
\end{array} \right).
\eeq
\end{definition}
\lem
\label{blockcrit}
Suppose that the cone $C(d_{2}) \subset Her(d_{2}) \subset M(d_{2})$ .
Then $X$ is $PSD(d_{1}) \otimes C(d_{2})$-separable
iff $\tilde{\phi}(X) \succeq 0$ (i.e. is positive semidefinite) for all
$C(d_{2})$-positive linear operators $\phi: M(d_{2}) \longrightarrow
M(d_{1})$.  \end{lemma}

The proof uses the following proposition, which generalizes the
duality between positive linear maps and separable states.  
\begin{proposition} \label{positive maps dual to separable}
For Hermitian matrices $M \in Block(d_1, d_2)$, the following are 
equivalent:\\
\noindent
1. $\tr M Z \ge 0$ for all $PSD(d_1) \otimes C(d_2)$-separable
matrices $Z$. 
\\ \noindent
2. $M^{ij} = \chi(e_i e_j^\dagger)$ for some $\chi$ such 
that $\chi(PSD(d_1)) \subseteq C(d_2)^*$.
\end{proposition}

\noindent
{\bf Proof of Proposition \ref{positive maps dual to separable}}
Item 1 of the proposition is equivalent to the same 
statement with $Z$ of the form
$x x^\dagger \otimes Y$, with $xx^\dagger$ a rank-one matrix in 
$PSD(d_1)$ and $Y \in C(d_2)$.  This is equivalent to:
\beqa \label{hiho}
\sum_{ij} x_i x_j^*\tr (M^{ij} Y ) = \tr \sum_{ij} x_i x_j^* 
M^{ij} Y \ge 0\;.
\eeqa
Since this holds for any $Y \in C(d_2)$, this says
$\sum_i x_i x_j^* M^{ij} \in C(d_2)^*$.
Define a linear map $T_M: M(d_1) 
\rightarrow M(d_2)$ via $T_M(e_i e_j^\dagger) = 
M^{ij}$.  Then for any $x \in \C^{d_1}$, 
$T(x x^\dagger) = \sum_{ij} x_i x_j^* 
T_M(e_i e_j^\dagger) =   \sum_{ij} x_i x_j^* 
M^{ij}\;$,  which 
we have just shown is in $C(d_2)^*$.  
Since $xx^\dagger$ generate $PSD(d_1)$, this is equivalent
to saying $T_M$ 
takes $PSD(d_1)$ into $C(d_2)^*$, so $T_m$ is the desired
$\chi$.  The other direction ($2 \Rightarrow 1$)
is similar (most of the steps
above were equivalences).
\QED

We remark that 
essentially the same argument establishes a similar statement
for {\em arbitrary} pairs of cones of Hermitian matrices, 
in which  $PSD(d_1)$ is replaced by a cone $C(d_1)$ in item 1, and
by $C(d_1)^*$ in item 2.

\noindent
{\bf Proof of Lemma 1:}  ``Only if'' is trivial:  $PSD(d_1) \otimes C(d_2)$-separability
of $X$ means $X = \sum_i A_i \otimes B_i$, with $A_i \in PSD(d_1), B_i \in
C(d_2)$, hence $\tilde{\phi}(X) = \sum_i A_i \otimes \phi(B_i)$; 
since we assumed  $\phi(C(d_2)) \subseteq PSD(d_1)$, we have $\phi(B_i) \in 
PSD(d_1)$, so $\tilde{\phi}(X) \in Sep(d_1, d_1)$.  

For ``if,'' note that
$\tilde{\phi}(X) \succeq 0$ 
says that for any positive semidefinite $B \in Block(N,K)$, 
\beqa \label{wombat}
\sum_{ij} \tr B^{ij} \phi(X^{ij}) \ge 0\;.
\eeqa
By the definition of dual cone (and the self-duality of $PSD(d_1)$) it is 
easily seen that 
$\phi^\dagger(PSD(d_1)) \subseteq C(d_2)^*$.
Now, (\ref{wombat}) is 
equivalent to 
\beqa
\sum_{ij} \tr \phi^\dagger(B^{ij}) X^{ij} 
= \tr \tilde{\phi^\dagger}(B) X
\ge 0\;.
\eeqa
Letting $B^{ij} = e_ie_j^\dagger$, so $B$ is 
the block matrix of a positive semidefinite rank-one state
(specifically, the  
unnormalized maximally entangled state $xx^\dagger$ with  
$x = \sum_i e_i \otimes e_i$), we have that the 
matrix
$\tilde{\phi^\dagger}(B)$ satisfies condition $2$ of the Proposition;  
as $\phi$ ranges over all $PSD(d_1)$-to-$C(d_2)^*$-positive 
maps, $\tilde{\phi^\dagger}(B)$ ranges over all such matrices
so by Proposition \ref{positive maps dual to separable}
$X$ is $PSD(d_1) \otimes C(d_2)$-separable.
\QED

The next proposition will allow us to extend the (exact)
bipartite result from \cite{GB02} to multiparty systems.  
The bipartite result was that everything in the ball 
$B(N,1) := \{I+\Delta: ||\Delta||_2 \le 1\}$ of 
size $1$ in Frobenius norm around the identity operator is separable;
therefore, so is everything in the cone, 
which we call $G(d_1 d_2, 1)$,  generated by
that ball, for a bipartite system with subsytems of 
dimensions $d_1,d_2$.  We define a slight generalization
of this cone:
\begin{definition}
Let $G(N,a) \subset Her(N) \subset M(N)$ be the cone generated by
hermitian $N \times N$ matrices of the form $\{I +\Delta :
||\Delta||_{2} =: (tr(\Delta\Delta^{\dagger})^{\frac{1}{2}} \leq a \}
$.
\end{definition}
A sufficient condition for tripartite
separability of $X$ is clearly that it belong to the cone
generated by $A_i \otimes Z_i$, where
$A_i \in PSD(d_1)$ and $Z_i \in G(d_2d_3, 1)$; this can be
used to derive a tripartite sufficient condition for 
separability in terms of Frobenius norm.  Letting this 
tripartite 2-norm ball generate a cone of separable tripartite
states, similar reasoning gives a ball of 4-partite states,
and so on.   The key to the induction step is 
the following proposition.

\pro
\label{contract}
 If $\phi: M(N) \longrightarrow M(K)$ is a $G(N,a)$-positive
linear operator (i.e. $\phi(X) \succeq 0$ for all $X \in G(N,a)$ and
$\phi(I) = I \in M(K)$ then
\begin{enumerate}
\item
$||\phi(X)||_{\infty} \leq a^{-1} ||X||_{2} $ for all hermitian $X \in M(N)$.
\item
$||\phi(Y)||_{\infty} \leq a^{-1} \sqrt{2} ||Y||_{2} $ for all $Y \in M(N)$.
\end{enumerate}
\epro 

\noindent
{\em Proof:}  All $G(N,a)$-positive $\phi$ satisfy
$\phi(I + \Delta) \succeq 0$ for all $\Delta$ such that 
$||\Delta||_2 \le a$, which when $\phi(I) = I$ gives 
$I + \phi(\Delta) \succeq 0$; by Proposition 
\ref{elementary, my dear Watson} this is equivalent to 
$||\phi(\Delta)||_\infty \le 1$, establishing item 1.  
Item 2. uses item 1 and the following:
\vskip 4pt
\noindent
If a linear operator $\phi: M(N) \longrightarrow M(K)$
satisfies $\phi(Her(N)) \subset Her(K)$ and $||\phi(Z)||_{\infty}
\leq ||X||_{2} $ for all hermitian $Z \in M(N)$ then
$||\phi(Y)||_{\infty} \leq \sqrt{2} ||Y||_{2} $ for all $Y \in
M(N)$.

To show this, let $Y = A + iB$ with $A, B$ Hermitian.
Now, $\phi(Y) = \phi(A) + i \phi(B)$, so 
$||\phi(Y)||_\infty \le ||\phi(A)||_\infty + ||\phi(B)||_\infty$.
This is less than or equal to $||A||_2 + ||B||_2$ by assumption
($A,B$ being Hermitian).  The conclusion follows by the square root of 
the elementary inequality $(x+y)^2 \le 2 (x^2 + y^2)$ 
(obtained from 
$2xy \le x^2 + y^2$, which comes from $(x-y)^2 \ge 0$).
\QED

The following example shows that the extra $\sqrt{2}$
factor is the best possible in this statement. 

Consider $\phi:
M(2) \longrightarrow M(2)$, $\phi(X) = X(1,1)A_{1} + X(2,2) A_{2}$;
where $A_{1}, A_{2} $ are real symmetric anticommuting unitary
matrices:
$$
 A_{1} = \left( \begin{array}{cc}                      
                  1& 0 \\
		  0 & -1 \end{array} \right) ,	 
$$
$$
 A_{2} = \left( \begin{array}{cc}                      
                  0& 1\\
		  1 & 0 \end{array} \right) .	 
$$
Notice that for real $a,b$ we have that $aA_{1} + b A_{2} = (a^{2} +
b^{2})^{\frac{1}{2}} U$ for some real symmetric unitary $U$. Thus
$\phi(Her(2)) \subset Her(2)$ and $||\phi(Z)||_{\infty} \leq ||Z||_{2}
$ for all hermitian $Z \in M(2)$.  Consider the following
(nonhermitian) matrix :
$$
 Y = \left( \begin{array}{cc}                      
                  1& 0 \\
		  0 & i \end{array} \right) .	 
$$
Then $||Y||_{2}^{2} = 2 $ and $||\phi(Y)||_{\infty}^{2} =
||\phi(Y)||_{2}^{2} = 4$, since $Det(\phi(Y)) = 0$.
\QED

\noindent
{\bf Problem: } Is the extra $\sqrt{2}$ factor still the best possible if,
additionally, $ \phi(I) = I $?

Proposition 2 of  \cite{GB02} is a similar contraction result
with constant $1$ rather than $\sqrt{2}$ on all matrices, not just
Hermitian ones, for the usual positive maps; the proofs used
there do not work for the different notion of positivity used here.

Now everything is ready for our attack on multipartite separability.

\thm 
\label{theorem: multipartite preliminary}
Let $H_1, H_2$ have dimensions
$n_1, n_2$.  
If an unnormalized density matrix $\rho: H_{1} \otimes
H_{2}  \longrightarrow H_{1}  \otimes H_{2}$ 
satisfies the inequality $||\rho-I||_{2} \leq a/\sqrt{2} $ then
it is $PSD(n_1) \otimes G(n_2, a)$-separable.  \ethm

\noindent
{\em Proof: }

Let $\rho = I + \Delta$ $\Delta$ Hermitian;  by Lemma $1$, 
we are looking for a bound on $||\Delta||_2$ that ensures, 
for any $G(n_2,a)$-to-$PSD(n_1)$-positive linear operator, that
$\tilde{\phi}(I + \Delta) \succeq 0$.  $\tphi(I) = I$, so 
$\tphi(I + \Delta) = I + \tphi(\Delta)$;  
$||\tphi(\Delta)||_\infty \le 1$
will ensure this (cf. Proposition 
\ref{elementary, my dear Watson}).  The argument establishing that 
$||\tphi(\Delta)||_\infty \le 1$ is essentially
identical to the proof of the main theorem of 
\cite{GB02}, except that because $\phi$ is not an ordinary
positive map 
we must use the weaker contraction bound of Proposition \ref{contract}, 
with its $\sqrt{2}$ factor, in place of \cite{GB02}'s result with 
a factor $1$.
\beqa
||\tphi(\Delta)||_\infty ^2 \le ||A||_\infty ^2 \le ||A||^2_2,
\eeqa
where $A := [a_{ij}]$, $a_{ij} := ||\phi(\Delta^{ij})||_\infty $.
\beqa
||A||_2^2 = \sum_{ij} a_{ij}^2 = \sum_{ij} ||\phi(\Delta^{ij})||_\infty ^2.
\eeqa
(The first inequality is because the operator norm of a block matrix
is bounded above by that of the matrix whose elements are the norms
of the blocks, and the second is because the Frobenius norm is an upper
bound to the operator norm.)
$||\phi(\Delta^{ij})||_\infty ^2 \le 2 a^{-2} ||\Delta^{ij}||_\infty ^2$ by Proposition
\ref{contract}, and it is an elementary norm inequality
that $||\Delta^{ij}||_\infty  \le ||\Delta^{ij}||_2$.  
So
\beqa
||\tphi(\Delta)||_\infty^{2}  \le 2 a^{-2}
\sum_{ij} ||\phi(\Delta^{ij})||_\infty ^2 \nonumber \\
\le 2 a^{-2} \sum_{ij} ||\Delta^{ij}||^2_2 
\equiv 2 a^{-2} ||\Delta||^2_2.  
\eeqa
Thus if $||\Delta||_2 \le a/\sqrt{2}$, $||\tphi(\Delta)||_\infty
\le 1$.
\QED

\begin{corollary}\label{cor: multipartite general}
If an $m$-partite unnormalized density matrix $\rho:
H_1 \otimes \cdots \otimes H_m 
\longrightarrow 
H_1 \otimes \cdots \otimes H_m$ satisifes $||\rho - I ||_2 
\le 1/(2^{m/2-1})$ then it is separable.
\end{corollary}

\noindent
{\bf Proof:}
The main result of \cite{GB02} is that  
$G(d_1d_2, 1)
\subset Sep(d_1,d_2)$.  This is the base case
for an induction on the number of subsystems.
For the induction step, fix $m > 2$ and 
suppose as our induction hypothesis the 
Corollary holds for $m-1$, i.e. 
$G(d_1,...,d_{m-1}, 2^{-((m-1)/2 -1)}) \subseteq 
Sep(d_1,...d_l)$.  
Theorem \ref{theorem: multipartite preliminary} tells us 
$G(d_1,...d_{m}, 2^{-((m-1)/2 -1)}/\sqrt{2} \equiv 
2^{-(m/2 -1)})$ is $PSD(d_m) \otimes 
G(d_1...d_{m-1}, 2^{-(m-1)/2 -1)})$-separable, and therefore 
(by the induction hypothesis and
the recursive definition of separability) separable.
\QED  

If we had an analogue to Theorem \ref{theorem: multipartite preliminary}, 
with $G(n_1, a)$ instead
of $PSD(n_1)$, and some constant $\alpha$ replacing $1/\sqrt{2}$, 
then we could get one over
a polynomial instead of an exponential in Corollary 
\ref{cor: multipartite general}, by recursively dividing
systems into subsystems of more or less equal size, since
this involves a logarithmic number of partitionings compared to
splitting off one system at a time.  
The first step toward such a theorem
would be to apply the characterization of 
$C(d_1) \otimes C(d_2)$-separability analogous to Proposition
\ref{positive maps dual to separable} (and discussed after that
proposition above) to $G(d_1, a), G(d_2, a);$ the fact that 
neither of these cones is self-dual has so far proved an obstacle
to our getting useful results along these lines.  

\thm Consider $\rho \in RLin(d_1,...,d_m)$. If $||\rho-I||_{2} \leq 1$
then \beq \rho = \sum a_{i} \rho^{(i)} \otimes \rho_{i}, a_{i} \geq 0
\eeq where for all $i$ $ \rho^{(i)}_1$ is a real positive semidefinite
$d_{1} \times d_{1}$ matrix; $\rho_{i} \in RLin(d_{2},...,d_m)$ and
$||I - \rho_{i}||_{2} \leq 1$.
\end{theorem}

\proof  The proof goes essentially like that of Theorem
\ref{theorem: multipartite preliminary}
except that the blocks 
$\Delta^{ij}$ are Hermitian by Proposition \ref{def: RLin}.  
Consequently we may use item 2. rather than item 1. of
Proposition \ref{contract}, and obtain the larger radius ball.
\QED

\cor If $\rho \in RLin(d_1,...,d_m)$ and $||\rho-I||_{2} \leq 1$ then
$\rho$ is real-separable.  In other words the maximal separable ball in
$RLin(d_1,...,d_m)$ around the identity $I$ has radius $1$ .  \ecor

The next Proposition is immediate from results of
\cite{GB02}, derived 
using ``scaling,''
i.e., considering all ways of writing a matrix
$\rho$ as a positive scalar times the sum of the identity and a
Hermitian perturbation, and minimizing the 2-norm of the 
perturbation).

\begin{proposition}
Define $\mu(\rho)$ as the maximum of $||\Delta||_2$ over
all $\Delta$ such that there exists an $\alpha > 0$ for 
which $\rho = \alpha(I + \Delta)$.
Let $\rho$ be a normalized ($\tr \rho = 1$) density matrix.
Then the following three statements are equivalent:\\
1. $\mu(\rho) \le a$.\\
2. $\tr \rho^2 \le 1/(d-a^2)$.\\
3. $||\rho - I/d||_2 \le a/\sqrt{d(d-a^2)}$. 
\end{proposition}

Using this Proposition, Theorem \ref{theorem: multipartite preliminary}
has (via Corollary \ref{cor: multipartite general}) the following
corollary:

\cor
If an $m$-partite normalized (i.e. unit trace) density matrix $\rho:
H_1 \otimes \cdots \otimes H_m 
\longrightarrow 
H_1 \otimes \cdots \otimes H_m$ satisifes $||\rho - I/d ||_2 
\le \frac{1}{2^{m/2-1 d}  }$,
where $d = dim ( H_1 \otimes \cdots \otimes H_m )$, then it is separable.
\ecor
(The proposition actually gives the (negligibly) 
tighter statement with 
$2^{m/2-1}\sqrt{d(d-2^{-(m-2)})}$ in the denominator.)

\section{Discussion}

In many interesting experimental or theoretical situations, the system
is in a ``pseudopure state'': a mixture of the uniform density matrix
with some pure state $\pi$: \beqa \label{pseudopure state}
\rho_{\epsilon, \pi} := \epsilon \pi + (1 - \epsilon) I/d\;, \eeqa
where $d = d_1,..d_m$ is the total dimension of the system.  For
example, consider nuclear magnetic resonance
(NMR) quantum information-processing (QIP), where $d=2$
(the Hilbert space of a nuclear spin), and $m$ is the number of
spins addressed in the molecule being used.  As discussed in
more detail below, the initialization procedures standard in most
NMRQIP implementations prepare pseudopure states.

Write 
\beqa
\rho_{\epsilon, \pi} = (1/d)I + \epsilon(\pi - I/d)\;.
\eeqa
Since $||\epsilon (\pi - I/d)||_2 = 
\epsilon \sqrt{\frac{d-1}{d}}$,  by Corollary 3, this is separable if 
\beqa
\epsilon \le 2^{-(m/2 - 1)}/ \sqrt{d(d -1)}\;,
\eeqa
For $m$ $D$-dimensional systems (so $d=D^m$), this implies
the (negligibly loosened) bound 
\beq \label{eq: earlier bound}
\epsilon \le 2^{-(m/2-1)}/D^m \;.
\eeq
This is an exponential improvement over the result in \cite{Rungta2001a}
(the qubit case is in \cite{Braunstein99a})
of $\epsilon \le 1/(1 + D^{2m-1})$.  For $m$ qubits, for example, 
our result goes asymptotically as $2^{-((3/2)m -1)}$, versus $2^{-(2m-1)}$  in \cite{Braunstein99a}.
Another comparison 
is with our earlier bound of 
\beq \label{bipartite bound}
\epsilon \le 1/(D^m - 1)
\eeq
guaranteeing separability 
for $m$ $D$-dimensional systems
with 
respect to every bipartition \cite{GB02}.   
It is interesting that this is exponentially
larger than the present bound guaranteeing multipartite separability,
although we do not know that a tight multipartite bound would still
exhibit this exponential separation.  

In
liquid-state NMR at high temperature $T$, 
the sample is placed in a high DC magnetic field.
Each spin is in a highly mixed thermal state.  It is not
maximally mixed because of the energy splitting between the higher-energy
state in 
which the spin is aligned with  
the magnetic field, and the higher-energy
one in which 
it is anti-aligned.   This gives probabilities for those 
states proportional to the 
Boltzmann factors $e^{\pm \beta \mu B}$, where $\beta \equiv 1/kT$ with 
$k$ Boltzmann's constant, $\mu$ the magnetic moment of the nuclear
spin, $B$ the external field strength.  For realistic high-T liquid
NMR values of $T=300$ Kelvin, $B = 11$ Tesla, $\beta \mu B \approx 
3.746 \times 10^{-5} \ll 1$.  Calling this $\eta$, 
$e^{\pm \eta} \approx 1 \pm \eta$, so the probabilities
are $p_{\uparrow} 
\approx (1 - \eta)/2$, 
$p_{\downarrow} 
\approx (1 + \eta)/2$,   
where $\uparrow$/$\downarrow$ 
denote alignment/anti-alignment of the spin with the
field.  
With independent, distinguishable nuclear
spins, Maxwell-Boltzmann statistics give 
the highest-probability pure state, 
with all $m$ spins up (field-aligned), probability about 
$(1 + \eta)^m/2^m \approx (1 + m \eta)/2^m$.
Standard pseudopure-state preparation creates a mixture of this most
probable pure state
and the maximally mixed state, by applying a randomly chosen
unitary 
from the group of unitaries fixing the all-spins-aligned state.
($U$ could be chosen uniformly (i.e. 
with Haar measure on this group), but efficient 
randomization procedures may draw from carefully chosen
finite sets of such unitaries \cite{Knill98a}.)
Thus, we get a mixture
\beqa
(1 - \epsilon) I/2^m + \epsilon \proj{\uparrow \cdots \uparrow}\;,
\eeqa
with 
\beqa \label{pseudopure polarization}
\epsilon = \eta m/2^m\;.
\eeqa
With 
$\eta \approx 3.746 \times 10^{-5}$.
this implies that below about 23 qubits, 
NMR pseudopure states are all separable, compared to 
the $\approx 13$ qubits one gets from the bound in 
\cite{Braunstein99a}.

Even without entanglement considerations, 
it is clear that pseudopure-state NMR quantum
computing will not give asymptotic gains over classical computing,
because of the exponentially increasing signal-to-noise
ratio from (\ref{pseudopure polarization}).  As pointed
out in \cite{Knill98b}, it does not follow that no application of 
liquid-state NMR can have better-than-classical asymptotic 
performance: NMRQIP is not limited to pseudopure-state initialization.
It is not known that NMRQIP with other initialization
schemes,
such as those that involve preparing a {\em fixed} number of pseudopure
qubits as the total number of spin qubits grows,
can be efficiently classically simulated,
and even with one pure qubit interesting things can be efficiently
done for which no efficient classical algorithm is currently
known \cite{Knill98b}.  Another non-pseudopure initialization scheme
is the Schulman-Vazirani algorithmic cooling procedure \cite{SV99}, which 
essentially uses an efficient (and NMR-implementable)
compression algorithm to convert 
the thermal state 
with  entropy $S$, for $m$ nuclear spins, into $\log{S}$ 
maximally mixed qubits and $m - \log{S}$ pure ones.  
Their work shows that the theoretical model derived from NMR
with an initial thermal state is as powerful as standard
quantum computation.
Though
the overhead required is polynomial, the space overhead
is large enough to be impractical given the small number
of qubits available in liquid-state NMR.  But 
algorithmic cooling is
certainly relevant in principle to the asymptotic
power of an implementation, and could be practically  
relevant
to a high temperature
bulk QIP implementation that was expected to be
sufficiently scalable that asymptotic considerations 
are relevant.  

However, there is still
the interesting possibility that one may
produce an entangled overall density matrix (and not just 
a mixture of the maximally mixed state with an entangled state)
via pseudopure-state NMRQIP.  The results herein increase the 
number of qubits known to be 
required before this may be possible, although, 
since we have not shown that the bounds herein are tight, with 
our assumed $\eta$ even
at 23 qubits there is no guarantee one can prepare an entangled 
pseudopure state.  By contrast, from (\ref{bipartite bound})
and (\ref{pseudopure polarization}) 
one needs $m = 1/\eta$ qubits (about $26,700$ for our $\eta$) 
to have any hope of obtaining a pseudopure state that
is not bipartite-separable with respect to a
partition of the qubits into two sets. Again, we remind the
reader that non-pseudopure protocols could conceivably
give such a ``biseparable'' state with far fewer qubits;
in the Appendix we discuss some implications of our results for
this possibility.

In conclusion we have derived an upper bound, exponentially better
than those already known, on the Frobenius-norm radius of a
ball of separable matrices around the identity matrix. 
The bound has implications for the minimum polarization needed for 
bulk quantum-information-processing protocols initialized by
preparing pseudopure states from a thermal state by via averaging,
to produce entanglement:  known lower bounds on this minimal 
polarization are exponentially increased by our results.  We stress,
however, that this
is just one application of a general, computationally simple 
sufficient criterion for multipartite
entanglement, applicable to states of any form.
Its geometric nature should make it useful in 
many applications, both theoretical and practical.  The 
question of whether this bound is tight, or whether there 
is a larger 
separable Frobenius norm ball around the identity, remains open.

\begin{appendix}
\section{Entanglement and thermal initial states in NMR}
Schulman and Vazirani's algorithmic cooling protocol
shows that it is, in theory, possible to prepare any entangled
state from sufficiently many thermal NMR qubits.  
The question of just how many qubits are required by means 
possibly simpler than algorithmic cooling is also of interest.
One can gain some information about this using our results, by
applying Corollary 3 to the initial thermal density matrix of an NMR 
system.  This matrix, which is approximately 
\beqa
\left( \begin{array}{cc}
		  \frac{1 + \eta}{2} & 0 \\
		  0 & \frac{1- \eta}{2}  
\end{array} \right)^{\otimes n}
\eeqa
(with each qubit expressed 
in the $\ket{\uparrow}, \ket{\downarrow}$ basis), has
\beqa
||\rho - I/d ||_2^2 = \frac{(1 + \eta^2)^m -1}{2^m} 
\approx m \eta^2/2^m\;.
\eeqa
This should be compared to Corollary 3's separability condition, 
which guarantees separability if this squared distance is below
$2^{-(3m-2)}$.  The comparison gives that for $m$ qubits,
the thermal state (and any state reachable from it by unitary 
transformation) is separable if
\beqa
\eta \le m^{-1/2} 2^{m-1}\;.
\eeqa  
For the same experimental conditions considered above, 
14 qubits are required before this bound is exceeded (rather than
the 23 for the pseudopure state prepared from this thermal
state).

If one wants the possibility of bipartite entanglement with 
respect to some partition of the qubits into two sets, it is
necessary to beat the bound (closely related to 
(\ref{bipartite bound})) 
\beqa
||\rho - I/d ||_2^2  \le \frac{1}{d(d-1)}\;.
\eeqa  
For the thermal state, this gives
$\eta \le m^{-1/2} 2^{-m/2}$. 
With $\eta = 3.746 \times 10^{-5}$ as before, the bound
is not surpassed until 25 qubits.  Although this comes from an 
essentially tight bound, that does not imply that entanglement can
be achieved through computation starting with this initial state.
Although the thermal state has the same magnitude perturbation
as some entangled state, the latter will have different eigenvalues,
so unitary manipulation will not get us there, and it is an 
interesting question what we can achieve along these lines
using 
NMR-implementable
nonunitary manipulations (which may sometimes require extra thermal ancilla 
bits that should be counted as resources).
To understand when {\em unitary} manipulations might be guaranteed
to give us entanglement, a promising approach is to look for 
conditions on the spectrum of a matrix sufficient for an entangled 
state with that spectrum to exist.  We conjecture that the problem
of determining, from a spectrum, whether or not an entangled state
with that spectrum exists, is NP-hard in terms of an appropriate
measure of problem size;  this does not rule out easier-to-evaluate
sufficient conditions, perhaps obtained from relaxations of the
above problem.  Theorem 4 of \cite{GB02} gives some information on 
spectra sufficient to guarantee entanglement.

For the pseudopure fraction $\epsilon$ there is an
{\em upper} bound of $2/(2 + 2^m)$ from \cite{GB02}
nearly matching the lower bound (\ref{bipartite bound}).  Comparison
to the expression 
$\epsilon = \eta m/2^m$ for pseudopure 
polarization suggests that, at some large
number of qubits $m \ge 2/\eta$, even pseudopure protocols will exceed this 
bound.    With 
$\eta = 3.746 \times 10^{-5}$ this gives about 53,400 qubits.
This is far beyond the range generally viewed as relevant for liquid state NMR,
and improved polarization is unlikely to bring it into this range.
Since the state demonstrating entanglement at the upper
bound is not in general 
pseudopure, exceeding this bound is still no guarantee we can get
entanglement.
Also, this is well into the range where
algorithmic cooling could produce much stronger entanglement.
Still, one can imagine that in other bulk QIP implementations the
balance between the difficulty of implementing complex unitaries
(relatively easy in NMR) and the difficulty of preparing large numbers
of thermal qubits (apparently relatively hard in NMR) could be
different, and entanglement generation by simple manipulations on 
thermal states, perhaps even by pseudopure state preparation,
might be promising.
\end{appendix}
 
\section*{Acknowledgments}
We thank Manny Knill and Isaac Chuang for discussions,
and the US DOE and NSA for financial support.


\begin{thebibliography}{8}
\expandafter\ifx\csname natexlab\endcsname\relax\def\natexlab#1{#1}\fi
\expandafter\ifx\csname bibnamefont\endcsname\relax
  \def\bibnamefont#1{#1}\fi
\expandafter\ifx\csname bibfnamefont\endcsname\relax
  \def\bibfnamefont#1{#1}\fi
\expandafter\ifx\csname citenamefont\endcsname\relax
  \def\citenamefont#1{#1}\fi
\expandafter\ifx\csname url\endcsname\relax
  \def\url#1{\texttt{#1}}\fi
\expandafter\ifx\csname urlprefix\endcsname\relax\def\urlprefix{URL }\fi
\providecommand{\bibinfo}[2]{#2}
\providecommand{\eprint}[2][]{\url{#2}}

\bibitem[{\citenamefont{Braunstein et~al.}(1999)\citenamefont{Braunstein,
  Caves, Jozsa, Linden, Popescu, and Schack}}]{Braunstein99a}
\bibinfo{author}{\bibfnamefont{S.}~\bibnamefont{Braunstein}},
  \bibinfo{author}{\bibfnamefont{C.~M.} \bibnamefont{Caves}},
  \bibinfo{author}{\bibfnamefont{R.}~\bibnamefont{Jozsa}},
  \bibinfo{author}{\bibfnamefont{N.}~\bibnamefont{Linden}},
  \bibinfo{author}{\bibfnamefont{S.}~\bibnamefont{Popescu}}, \bibnamefont{and}
  \bibinfo{author}{\bibfnamefont{R.}~\bibnamefont{Schack}},
  \bibinfo{journal}{Physical Review Letters} \textbf{\bibinfo{volume}{83}},
  \bibinfo{pages}{1054} (\bibinfo{year}{1999}).

\bibitem[{\citenamefont{Rungta et~al.}(2001)\citenamefont{Rungta, Munro,
  Nemoto, Deuar, Milburn, and Caves}}]{Rungta2001a}
\bibinfo{author}{\bibfnamefont{P.}~\bibnamefont{Rungta}},
  \bibinfo{author}{\bibfnamefont{W.~J.} \bibnamefont{Munro}},
  \bibinfo{author}{\bibfnamefont{K.}~\bibnamefont{Nemoto}},
  \bibinfo{author}{\bibfnamefont{P.}~\bibnamefont{Deuar}},
  \bibinfo{author}{\bibfnamefont{G.~J.} \bibnamefont{Milburn}},
  \bibnamefont{and} \bibinfo{author}{\bibfnamefont{C.~M.} \bibnamefont{Caves}},
  in \emph{\bibinfo{booktitle}{Directions in Quantum Optics: A Collection of
  Papers Dedicated to the Memory of Dan Walls}}, edited by
  \bibinfo{editor}{\bibfnamefont{D.}~\bibnamefont{Walls}},
  \bibinfo{editor}{\bibfnamefont{R.}~\bibnamefont{Glauber}},
  \bibinfo{editor}{\bibfnamefont{M.}~\bibnamefont{Scully}}, \bibnamefont{and}
  \bibinfo{editor}{\bibfnamefont{H.}~\bibnamefont{Carmichael}}
  (\bibinfo{publisher}{Springer}, \bibinfo{address}{New York and Berlin},
  \bibinfo{year}{2001}), \bibinfo{note}{also arXiv.org e-print
  quant-ph/0001075}.

\bibitem[{\citenamefont{Gurvits and Barnum}(2002)}]{GB02}
\bibinfo{author}{\bibfnamefont{L.}~\bibnamefont{Gurvits}} \bibnamefont{and}
  \bibinfo{author}{\bibfnamefont{H.}~\bibnamefont{Barnum}},
  \bibinfo{journal}{Physical Review A} \textbf{\bibinfo{volume}{66}},
  \bibinfo{pages}{062311} (\bibinfo{year}{2002}).

\bibitem[{\citenamefont{Woronowicz}(1976)}]{Woronowicz76a}
\bibinfo{author}{\bibfnamefont{S.~L.} \bibnamefont{Woronowicz}},
  \bibinfo{journal}{Reports on Mathematical Physics}
  \textbf{\bibinfo{volume}{10}}, \bibinfo{pages}{165} (\bibinfo{year}{1976}).

\bibitem[{\citenamefont{Horodecki et~al.}(1996)\citenamefont{Horodecki,
  Horodecki, and Horodecki}}]{Horodecki96a}
\bibinfo{author}{\bibfnamefont{M.}~\bibnamefont{Horodecki}},
  \bibinfo{author}{\bibfnamefont{P.}~\bibnamefont{Horodecki}},
  \bibnamefont{and}
  \bibinfo{author}{\bibfnamefont{R.}~\bibnamefont{Horodecki}},
  \bibinfo{journal}{Phys. Lett. A} \textbf{\bibinfo{volume}{223}},
  \bibinfo{pages}{1} (\bibinfo{year}{1996}), \bibinfo{note}{(also arXiv.org
  e-print quant-ph/9605038)}.

\bibitem[{\citenamefont{Knill et~al.}(1998)\citenamefont{Knill, Chuang, and
  Laflamme}}]{Knill98a}
\bibinfo{author}{\bibfnamefont{E.}~\bibnamefont{Knill}},
  \bibinfo{author}{\bibfnamefont{I.}~\bibnamefont{Chuang}}, \bibnamefont{and}
  \bibinfo{author}{\bibfnamefont{R.}~\bibnamefont{Laflamme}},
  \bibinfo{journal}{Physical Review A} \textbf{\bibinfo{volume}{57}},
  \bibinfo{pages}{3348} (\bibinfo{year}{1998}).

\bibitem[{\citenamefont{Knill and Laflamme}(1998)}]{Knill98b}
\bibinfo{author}{\bibfnamefont{E.}~\bibnamefont{Knill}} \bibnamefont{and}
  \bibinfo{author}{\bibfnamefont{R.}~\bibnamefont{Laflamme}},
  \bibinfo{journal}{Physical Review Letters} \textbf{\bibinfo{volume}{81}},
  \bibinfo{pages}{5672} (\bibinfo{year}{1998}).

\bibitem[{\citenamefont{Schulman and Vazirani}(1999)}]{SV99}
\bibinfo{author}{\bibfnamefont{L.~J.} \bibnamefont{Schulman}} \bibnamefont{and}
  \bibinfo{author}{\bibfnamefont{U.}~\bibnamefont{Vazirani}},
  \bibinfo{journal}{Proceedings of the 31st Annual {ACM} Symposium on the
  Theory of Computing (STOC)} pp. \bibinfo{pages}{322--329}
  (\bibinfo{year}{1999}), \bibinfo{note}{earlier version is quant-ph/9804060}.

\end{thebibliography}



\end{document}